%% file: main.tex
\documentclass[conference]{IEEEtran}

\usepackage[utf8]{inputenc}   
\usepackage{newtxmath}        

\usepackage{amsmath}
\usepackage{esvect}           

\usepackage{array}
\usepackage{multirow}
\usepackage{booktabs}
\usepackage{tabularx}
\usepackage{makecell}
\usepackage{nicematrix}
\usepackage{colortbl}

\usepackage{graphicx}
\usepackage{wrapfig}
\usepackage{caption}
\usepackage{subcaption}

\usepackage{algorithm}
\usepackage{algpseudocode}

\usepackage{anyfontsize}
\usepackage[all]{nowidow}     
\usepackage[normalem]{ulem}   
\usepackage{xcolor}
\usepackage{enumitem}
\usepackage{balance}          

\usepackage{url}
\usepackage{hyperref}
\usepackage{orcidlink}

\usepackage[most]{tcolorbox}
\usepackage[framemethod=TikZ]{mdframed}
\usepackage{framed}

\definecolor{light-gray}{gray}{0.9}
\hypersetup{
    colorlinks=true,
    linkcolor=black,
    filecolor=magenta,      
    urlcolor=blue,
    citecolor=black,
    pdfborder={0 0 0}
}

\newcolumntype{Y}{>{\centering\arraybackslash}X}

\definecolor{commentGray}{RGB}{120,120,120}
\renewcommand{\algorithmiccomment}[1]{\bgroup\color{commentGray}{//#1}\egroup}

\usepackage{pifont}
\usepackage{listings}
\usepackage{framed}
\usepackage{tikz}
\definecolor{light-gray}{gray}{0.9}
\usepackage[framemethod=TikZ]{mdframed}
\mdfsetup{skipabove=5pt,skipbelow=3pt}
\usepackage{lipsum}
\mdfdefinestyle{RQFrame}{%
	linecolor=black,
	outerlinewidth=0.15pt,
	roundcorner=3pt,
	innertopmargin=2pt,
	innerbottommargin=2pt,
	innerrightmargin=4pt,
	innerleftmargin=4pt,
	backgroundcolor=light-gray}

\definecolor{javagreen}{rgb}{0.25,0.5,0.35} 

\lstdefinestyle{Alg}{
  basicstyle=\ttfamily\footnotesize,
  breaklines=true,
  tabsize=2,
  mathescape,
  numbers=left,
  xleftmargin=2.5em,
  xrightmargin=0.5em,
  frame=tb,
  framexleftmargin=2em,
  emph={Algorithm,Input,Output,for,each,do,if,else,Function,while,let,be,repeat,until,return,times,and,or,break,in,then,},
  emphstyle={\textbf},
  escapechar=?,
  morecomment=[l][\color{javagreen}]{//},
  columns=flexible,
}

\begin{document}

\newcommand{\approach}[0]{QAMR}
\newcommand{\llm}[0]{{\text{\fontsize{8.5}{8.5}\textsf{LLM-Core}}}}
\newcommand{\llmEval}[0]{{\text{\fontsize{8.5}{8.5}\textsf{LLM-Eval}}}}

\title{Question Answering  for Multi-Release Systems: \\ A Case Study at Ciena}

\author{
    \IEEEauthorblockN{%
    Parham Khamsepour\IEEEauthorrefmark{1}\IEEEauthorrefmark{2}, 
    Mark Cole\IEEEauthorrefmark{2},
    Ish Ashraf\IEEEauthorrefmark{2},
    Sandeep Puri\IEEEauthorrefmark{2},
    Mehrdad Sabetzadeh\IEEEauthorrefmark{1},
    Shiva Nejati\IEEEauthorrefmark{1}
   }
       \IEEEauthorblockA{\IEEEauthorrefmark{1}University of Ottawa, 800 King Edward Avenue, Ottawa ON K1N 6N5, Canada}

    \IEEEauthorblockA{\IEEEauthorrefmark{2}Ciena Corp, 7035 Ridge Road, Hanover, MD 21076, USA
    }
	Email: \{parham.khamsepour,  m.sabetzadeh, snejati\}@uottawa.ca; \{mcole, iashraf,spuri\}@ciena.com
}

\maketitle

\begin{abstract}
Companies regularly have to contend with multi-release systems, where several versions of the same software are in operation simultaneously. Question answering over documents from multi-release systems poses challenges because different releases have distinct yet overlapping documentation. Motivated by the observed inaccuracy of state-of-the-art question-answering techniques on multi-release system documents, we propose \approach, a chatbot designed to answer questions across multi-release system documentation. \approach\ enhances traditional retrieval-augmented generation (RAG) to ensure accuracy in the face of highly similar yet distinct documentation for different releases. It achieves this through a novel combination of pre-processing, query rewriting, and context selection. In addition, \approach\ employs a dual-chunking strategy to enable separately tuned chunk sizes for retrieval and answer generation, improving overall question-answering accuracy. 

We evaluate \approach\ using a public software-engineering benchmark as well as a collection of real-world, multi-release system documents from our industry partner, Ciena. Our evaluation yields five main findings: (1)~\approach\ outperforms a baseline RAG-based chatbot, achieving an average answer correctness of 88.5\% and an average retrieval accuracy of 90\%, which correspond to improvements of 16.5\% and 12\%, respectively. (2)~An ablation study shows that \approach's mechanisms for handling multi-release documents directly improve answer accuracy. (3)~Compared to its component-ablated variants, \approach\ achieves a 19.6\% average gain in answer correctness and a 14.0\% average gain in retrieval accuracy over the best ablation. (4)~\approach\ reduces response time by 8\% on average relative to the baseline.
(5)~The automatically computed accuracy metrics used in our evaluation strongly correlate with expert human assessments, validating the reliability of our methodology.

\end{abstract}


\input{Sections/introduction}

\input{Sections/industry}

\input{Sections/approach}

\input{Sections/evaluation}

\input{Sections/related}

\input{Sections/lessons}
\input{Sections/con}

\newpage

\balance
\bibliographystyle{IEEEtran}
\bibliography{IEEEabrv,bib/ref}
\end{document}

%% file: Sections/introduction.tex
\section{Introduction}
\label{sec:intro}
Systems and software engineering processes rely on technical documents, such as requirements specifications, design documents, and installation guides. These documents support diverse stakeholders, such as end-users, product managers, software engineers, system architects,  and customer support teams, in understanding system functions, dependencies, and carrying out their roles.

Companies often maintain multiple versions of their systems, with each version evolving from its predecessors to meet client needs, support legacy users, comply with regulations, or enable phased rollouts~\cite{DBLP:conf/ewspt/Lehman96a}. Such collections of coexisting versions are referred to as \emph{multi-release systems}. Unlike a software product line~\cite{DBLP:books/daglib/0015277} -- a planned family of related products that coexist without forming sequential stages of evolution -- multi-release systems progress version by version. Each version typically has its own documentation, capturing updates, improvements, and configuration changes specific to that release while retaining much of the earlier content.

Large Language Models (LLMs) are now commonly used to develop question-answering chatbots in software and systems engineering. These chatbots help stakeholders search large document sets and get real-time answers to  questions. Among others, Abdellatif et al.~\cite{DBLP:journals/tse/AbdellatifBCS22,10.1145/3379597.3387472}, Daniel and Cabot~\cite{DBLP:journals/scp/DanielC24}, Chaudhary et al.~\cite{chaudhary2024developingllamabasedchatbotcicd}, and Huang et al.~\cite{iKnow} examine the use of chatbots in software engineering, both during development and operation, showing their potential to enhance communication and support tasks such as bug tracking, documentation, code generation, and runtime assistance.

One of the most widely used architectures for modern chatbots is retrieval-augmented generation (RAG). RAG integrates document retrieval with LLMs to produce more accurate and contextually relevant responses~\cite{10.5555/3495724.3496517,pmlr-v119-guu20a, DBLP:journals/corr/abs-2007-01282,10.1145/3644815.3644945}.
Applying RAG to documents associated with multiple releases of the same system presents two challenges: \textbf{(C1)} RAG's retrieval accuracy declines due to the coexistence of numerous versions with distinct yet overlapping content. For example, when a query pertains to a system property that varies across releases, the retrieval component of RAG may get confused, extracting inconsistent information. \textbf{(C2)} Standard RAG cannot effectively prioritize the most relevant information in the context of multi-release systems. Here, not all documents or their parts are necessarily linked to specific releases. To generate accurate answers, a query needs to be reinterpreted into more specific ones -- some tied to releases, and others independent of releases. This reinterpretation results in multiple sets of texts to be retrieved by RAG, but not all of these texts are equally relevant. A strategy is therefore needed to keep only the relevant texts retrieved by the different query~interpretations.

In this paper, we propose \approach, a novel RAG-based \textbf{Q}uestion \textbf{A}nswering chatbot for \textbf{M}ulti-\textbf{R}elease systems that addresses the above challenges. 
Our approach includes a pre-processing step in which documents are indexed into a separate corpus for each release, thereby preventing RAG from mixing information from different releases. Using a prompting technique, \approach\ transforms a given user query into multiple standalone versions that fully incorporate the conversational history. 
Unlike traditional RAG, which retrieves a single set of texts, \approach\ retrieves multiple sets, each using a rewritten standalone query. 
To use only the most relevant texts for answering the original user query, \approach\ employs a state-of-the-art LLM-based ranking strategy, RankRAG~\cite{yu2024rankragunifyingcontextranking}. Instead of using all retrieved documents for answer generation, RankRAG uses an LLM to rank and select the texts most contextually relevant to the input query before answer generation.

In addition to supporting multi-release systems, \approach\ mitigates a known limitation of RAG in  software-engineering applications~\cite{10.1145/3644815.3644945,chaudhary2024developingllamabasedchatbotcicd}: the sensitivity to chunk size  -- that is, the size of the retrieved texts. Small chunks often improve retrieval precision but provide limited context for answer generation, whereas large chunks supply more context but tend to lower retrieval accuracy. Instead of relying on a single compromise size, \approach\ employs a \emph{dual-chunking} strategy,  using smaller chunks \hbox{for retrieval and larger ones for answer generation.}

\par\noindent\textbf{Contributions.} \textbf{(1)} We introduce, \approach,  a RAG-based chatbot for question answering over multi-release systems. \approach\  can process structured content such as tables and schematics, and it generates multiple interpretations of user queries to improve retrieval accuracy. It employs RankRAG to identify contextually relevant information for response generation. In addition, \approach\ implements a dual-chunking strategy: it uses smaller text chunks during retrieval and larger chunks during answer generation. This design avoids the trade-offs inherent in traditional RAG systems that rely on a single chunk size.

\textbf{(2)} We evaluate \approach\ on an industry dataset from Ciena and on an open-source requirements set, REQuestA~\cite{ezzini2023aibasedquestionansweringassistance}. We compare \approach\ against a state-of-the-art baseline and conduct an ablation study to isolate the effect of each component of \approach. To do so, we report six \emph{LLM-as-Judge} metrics focused on answer correctness and retrieval accuracy~\cite{deepeval,liu2023gevalnlgevaluationusing}, along with response time.
Our use of LLM-as-Judge metrics is motivated by the limitations of both manual evaluation and traditional word-overlap metrics such as BLEU~\cite{Papineni02bleu:a}, ROUGE~\cite{lin-2004-rouge}, and cosine similarity:  Manual evaluation, while capable of nuanced judgment, is difficult to scale because it requires checking a large number of answers generated by a proposed chatbot, its variants, and  baselines, as well as the results of repeated runs to account for randomness. 
Automated metrics such as BLEU, ROUGE, and cosine similarity provide only shallow word-level matches and fail to capture the human-like qualities required to effectively evaluate LLM-generated texts; prior studies already show the inadequacy of these metrics in this regard~\cite{chaudhary2024developingllamabasedchatbotcicd,goyal2023newssummarizationevaluationera,liu2017evaluatedialoguesystemempirical,DBLP:conf/nlpcc/WuGSLJ23}. 

We implement two  measures to ensure that LLM-computed metrics serve as reliable proxies for expert human judgment. First, our LLM-as-Judge framework, created in collaboration with Ciena's subject-matter experts, uses prompts that closely reflect human evaluators' reasoning when assessing response quality in multi-release systems. 
Second, we validate the LLM-generated metrics against evaluations by Ciena's domain experts, who manually assess the accuracy of responses from an individual run of \approach\ on the full industrial dataset. Our analysis shows a strong correlation  between the LLM-generated metrics and expert assessments, confirming the reliability of our metrics.

\par\noindent\textbf{Findings.} Our evaluation results  show that \approach\ outperforms a baseline RAG-based chatbot without dual-chunking, achieving an average of 88.5\% answer correctness and 90\% retrieval accuracy.  These represent average improvements of 16.5\% and 12\%, respectively, over the baseline.

Our ablation study leads to two main findings: \emph{First,} the components of \approach\ complement one another. While each component individually improves accuracy, only the full \approach\ -- with all components included -- consistently achieves the strongest performance, yielding average gains of 19.6\% in answer correctness and 14.0\% in retrieval accuracy compared to the best ablation. \emph{Second,} the ability to reinterpret user queries across multiple system releases is critical for effective question answering on multi-release documents, as the ablation with this capability achieves higher answer correctness and contextual recall compared to the ablations without it.

With respect to response time, across the two datasets Ciena and REQuestA, \approach\ reduces the average by about 9\% and 8\%, respectively, yielding an overall improvement of roughly 8\% compared with the baseline.

\par\noindent\textbf{Novelty.} The novelty of our work is in designing a RAG-based chatbot for multi-release systems, addressing challenges from distinct but overlapping documentation. In addition, we propose a chunking strategy to mitigate the trade-off between retrieval accuracy and contextual richness by combining small chunks for retrieval with larger chunks for answer generation.

\par\noindent\textbf{Significance.} Multi-release systems are prevalent in industry, as organizations often have to deal with evolving products that have multiple active versions. A practical and accurate chatbot can provide substantial assistance in efficiently gleaning information of interest from the complex technical documents associated with such systems. Another significant aspect is our dual-chunking method, which has the potential to increase chatbot accuracy in broader contexts, meriting further study beyond multi-release systems.

\par\noindent\textbf{Replication Package.} All code, evaluation scripts, and experimental data for our public dataset are available online~\cite{replicationPackage}

%% file: Sections/industry.tex
\section{Industry Context and Motivation}
Ciena is exploring chatbots to automate routine tasks, with a focus on managing its complex multi-release system portfolio. Each release at Ciena requires updating  version-specific documentation that covers requirements, architecture, configuration, and  installation. While releases share substantial overlaps, they also include important differences.  These documents support diverse stakeholders -- engineers, developers, and product managers -- who each need to query documentation across single or multiple  releases. For instance, a software developer might ask a question concerning a specific release of a product, e.g., ``\emph{How do I update \textcolor{red}{[product]} to release 17?}''. To answer this question accurately, the information must be retrieved from documents corresponding to release 17. Similarly, the process for ``\emph{What are the upgrade paths to \textcolor{red}{[product]} release 12?}'' would require consulting the documents for release 12. The similarity in release-specific questions and the overlaps in the documentation of different releases make it essential to query the appropriate documents to generate correct responses. Alternatively, the developer could ask a generic question whose answer is not tied to a particular release, e.g., ``\emph{How to connect to \textcolor{red}{[product]}'s dashboard?}''. In such a case, Ciena would want to maintain the flexibility to provide a version-agnostic response that remains relevant across multiple releases. In all queries and examples, original content has been slightly modified for confidentiality, with redactions shown in square brackets (\textcolor{red}{[...]}).

Our goal in this paper is to develop a chatbot using state-of-the-art LLM technologies, enabling stakeholders at Ciena to query structurally complex and semantically rich system documentation while accommodating the multi-release nature of the company's systems.
While our work is motivated by the specific needs of Ciena, the challenges we address are not unique to the company. The prevalence of multi-release systems across industries necessitates the development of methods capable of handling the complexities of constantly evolving software documentation. Recognizing this broader relevance, we have designed our chatbot with a focus on generalizability. 

%% file: Sections/approach.tex
\section{Chatbot for Multi-Release Systems}
\label{section:entire-chatbot}
\approach\ takes multi-release documents and a query as input and generates a response based on the documents. Section~\ref{section:corpus-creation} describes corpus creation, and Section~\ref{section:architecture} presents \approach.

\subsection{Corpus Creation}
\label{section:corpus-creation}
\begin{figure}[t]
	\centering
	\includegraphics[width=0.85\columnwidth]{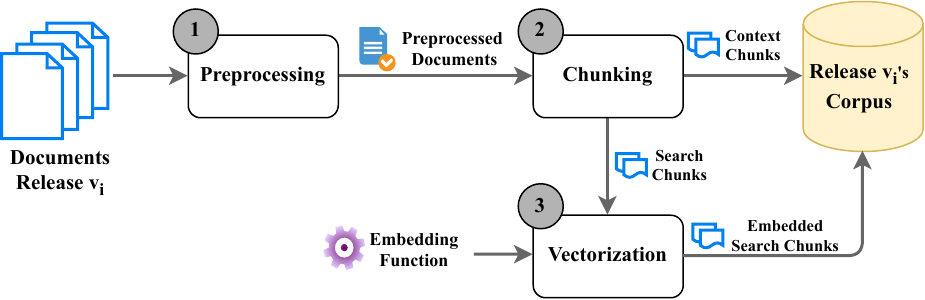} 
	\caption{\approach's corpus creation for each document release}
	\label{figure:corpus_creation}

\end{figure}

Given a set of multi-release documents, \approach\ creates a separate corpus for each release, instead of combining all releases into a single corpus. This approach ensures that differences between releases do not affect the LLM's accuracy during answer generation. Based on the user query, \approach\ selects the most relevant corpus for document retrieval (see Section~\ref{section:query-rewriting}). To transform the documents from a specific release into a corpus, we follow the three steps shown in Figure~\ref{figure:corpus_creation} and described next.

\begin{figure}[t]
	\centering
    \vspace*{-.5cm}
	\includegraphics[width=0.75\columnwidth]{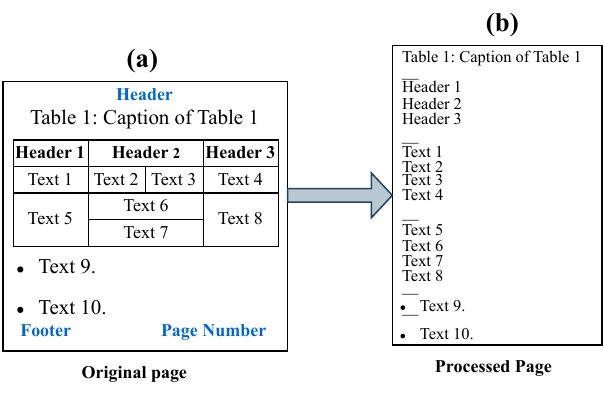}   
	\caption{Preprocessing of an individual page by \approach}
	\label{figure:document_processing}
\end{figure}

\textit{\textbf{Step 1: Preprocessing.}} 
To prepare documents for LLM processing, we extract text from all components, including tables, schematics, and flowcharts, in HTML or SVG formats.
Redundant elements, such as headers, footers, and copyright information, are removed to prevent \approach\ from retrieving irrelevant pages due to matching terms. Document names are added to each page's metadata. While not required for generating answers, this metadata ensures traceability between \approach's responses and the original documents. Text from tables, diagrams, and schematics is extracted from top-left to bottom-right, reconstructing lines to preserve both horizontal and vertical order. Delimiters are used to separate different sections. Figure~\ref{figure:document_processing} shows our preprocessing: headers, footers, and page numbers are removed; table text is captured top-left to bottom-right, converted to lines, and separated by ``---''.

\textit{\textbf{Step 2: Chunking.}} The processed text from Step~1 is divided into smaller segments, known as \emph{chunks}~\cite{10.1145/3644815.3644945,10.5555/3495724.3496517,chaudhary2024developingllamabasedchatbotcicd}. Chunking serves more than just fitting text into an LLM's token limit: For search, smaller chunks tend to increase retrieval accuracy. And, for answer generation, opting for the largest possible chunks (i.e., the LLM's token limit) is not necessarily the best option due to the ``needle in a haystack'' problem~\cite{DBLP:conf/emnlp/LabanFXW24,hassani2024rethinkinglegalcomplianceautomation, liu2023lostmiddlelanguagemodels,10.1145/3644815.3644945}.
To account for these considerations, we \emph{decouple} the search and answer-generation chunk sizes, allowing these sizes to be configured explicitly and independently. We generate two distinct sets of chunks: one for search and another for answer generation. We refer to this method, detailed in  Algorithm~\ref{alg:dual-chunking},  as \emph{dual chunking}.

\input{Sections/alg}

Algorithm~\ref{alg:dual-chunking} takes as input a set of processed documents, denoted as $\mathit{Docs}$, from Step~1. For each document in $\mathit{Docs}$, the algorithm produces two outputs: a set of \textit{search chunks} (\hbox{\textit{S-Chunks}}) and a set of \textit{context chunks} (\textit{C-Chunks}). The (smaller) search chunks are used for semantic similarity searches to accurately locate relevant text. After identifying the most relevant search chunks, the corresponding (and larger) context chunk is retrieved for each search chunk to generate the answer. Duplicates among context chunks are removed.

Algorithm~\ref{alg:dual-chunking} has two configurable parameters: $k$, the number of search chunks to create per page, and $\mathit{ps}$, the padding size for building context chunks. To create search chunks, each document page is divided into $k$ chunks, which are then added to \textit{S-Chunks} (lines~3–6). Each context chunk is created by extending a page with $\mathit{ps}$ characters from the bottom of its previous page at the beginning and $\mathit{ps}$ characters from the top of its next page at the end (lines~8–10). Thus, there is a one-to-one correspondence between context chunks and pages. To maintain traceability, the metadata of each context chunk is set to the title information of its corresponding document (line~11). Finally, the created context chunk is added to \hbox{\textit{C-Chunks}} (line~12). Once all the documents in \textit{Docs} have been processed, \textit{S-Chunks} and \textit{C-Chunks} are returned (line~15).

\textit{\textbf{Step 3: Vectorization.}} We generate embedding vectors for each search chunk using an embedding function. The resulting vectors numerically represent the chunks in a high-dimensional space, capturing their semantics~\cite{DBLP:conf/nips/MikolovSCCD13}. These vectors are stored in a vector database. We note that context chunks do not require vectorization in our approach. As shown in Algorithm~\ref{alg:dual-chunking}, there is a traceable relationship between search chunks and context chunks. Specifically, a search chunk $\mathit{sc}_{ijl}$ maps to its corresponding page $\mathit{pg}_{ij}$, which, in turn, maps to the associated context chunk $\mathit{cc}_{ij}$. Thus, when a search chunk is deemed relevant to a query (via semantic similarity), its corresponding context chunk can be directly identified.

\subsection{Chatbot Architecture}
\label{section:architecture}

\begin{figure}[t]
	\centering
	\includegraphics[width=0.9\columnwidth]{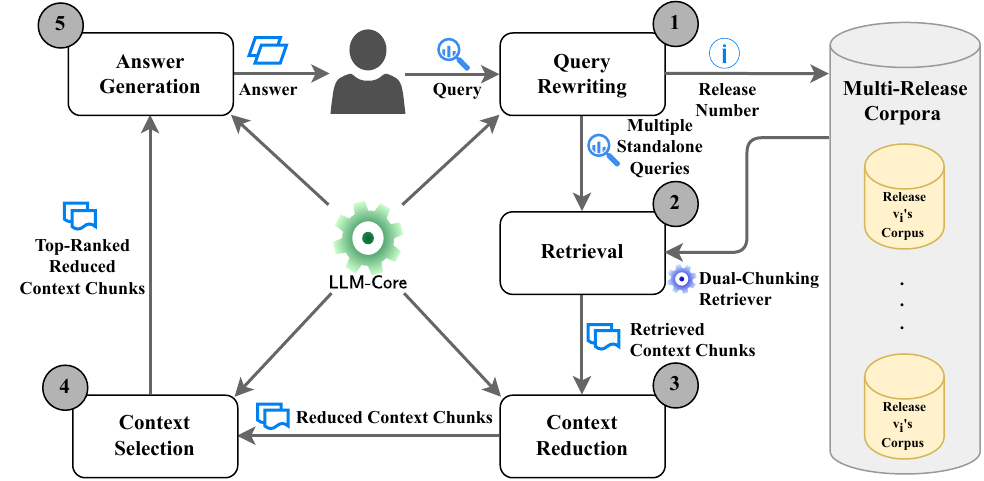} 
	\caption{Architecture of \approach.}
	\label{figure:system_architecture}
    \vspace*{-.2cm}
\end{figure}

Figure~\ref{figure:system_architecture} provides an overview of \approach, which consists of five steps for generating a response to a user query based on the corpora created in Section~\ref{section:corpus-creation}.  Steps 1, 3, 4, and 5 in Figure~\ref{figure:system_architecture} require  an LLM. While different LLMs could  be used for each step, we use a single LLM instance for efficiency, given the sequential nature of the steps. This instance is referred to as \llm\ in the figure and throughout the rest of the paper.
A working example~\cite{qamrexample} and the prompts of \approach~\cite{qamrprompts} -- omitted here for space -- are available online.

\textit{\textbf{Step 1: Query Rewriting.}}
\label{section:query-rewriting}
\approach\ begins by extracting the release number, if present, from the user query. The extracted release number is used in both the query rewriting process and for determining the appropriate corpus for document retrieval in Step~2. To extract the release number, \approach\ prompts \llm. This enables us to correctly map variations in the user query to the intended release. For instance, variants like ``R17.2'', ``Rel 17.20'', or ``Release 17.2'' are all interpreted to map to ``Release 17.20'', as it appears in the documents. Using an LLM for this mapping provides greater accuracy compared to regular expressions which are limited by fixed patterns.

After extracting the release number (if present), \approach\ rewrites the query using the user query, the conversation history, and the extracted release number. \approach\ generates three \emph{standalone} queries if a release number is identified; otherwise, it generates two.
A standalone query refers to a query that is self-contained and does not require additional context to be understood. Query rewriting is performed using three prompts that are available online~\cite{qamrprompts}. We refer to the three generated standalone queries as \emph{Base}, \emph{Filtered}, and \emph{Versionless}, as explained below.

\begin{figure}
	\centering
	\includegraphics[width=0.85\columnwidth]{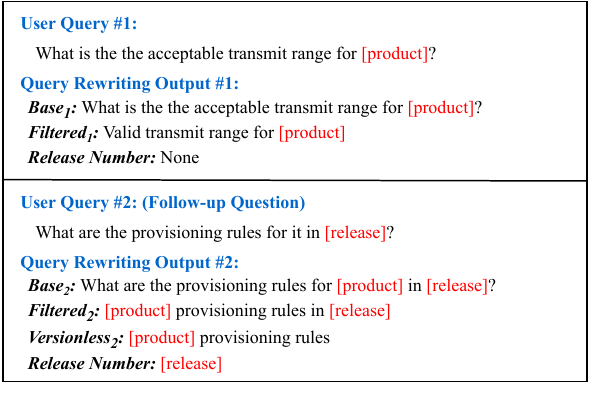} 
	\caption{Examples of query rewriting in \approach.}
	\label{figure:query_rewriting_example}
    \vspace*{-0.2cm}
\end{figure}

\textbf{The \emph{Base} query} interprets the user query in the context of the user conversational history. In interactive settings, users often ask questions that may lack clarity, particularly follow-up questions that rely on prior context, using pronouns like ``it'' or ``they'', or phrases such as ``the previous one''~\cite{ma2023queryrewritingretrievalaugmentedlarge,chaudhary2024developingllamabasedchatbotcicd,10.1145/3644815.3644945}.
Figure~\ref{figure:query_rewriting_example} illustrates two examples of \emph{Base} queries. The first example, \emph{Base}$_1$, is identical to the user query, as the query is already self-contained. In contrast, the second example, \emph{Base}$_2$, replaces the pronoun ``it'' with \textcolor{red}{[product]} \hbox{from the conversation history.}

\textbf{The \emph{Filtered} query} removes stop words and other common terms from the \emph{Base} query (above), focusing on core elements. The resulting query often aligns more closely with the content and language used in document tables, thereby increasing accuracy, specially when working with table-heavy documents. For example, in Figure~\ref{figure:query_rewriting_example}, \emph{Filtered}$_1$ and \emph{Filtered}$_2$ represent the \emph{Filtered} queries for their respective user~queries.

\textbf{The \emph{Versionless} query} is created -- only if the user query includes a release number -- by removing the release number from the \emph{Filtered} query (above). \emph{Versionless} queries address two scenarios that can cause retrieval inaccuracies: (1) when the release number is explicitly mentioned in relevant documents but absent in the user query, or (2) when the release number is explicitly included in the user query but missing from relevant documents. For example, in Figure~\ref{figure:query_rewriting_example}, \emph{Versionless}$_2$, formed by omitting the release number from \emph{Filtered}$_2$, is the third standalone query generated for the second user query. In contrast, for the first user query, which lacks a release number, only two standalone queries are generated.

\textit{\textbf{Step 2: Retrieval.}}
\label{section:doc-retrieval} 
In this step, we first retrieve the most relevant search chunks by taking the union of the highest-scoring (search) chunks that match a given standalone query, using two widely used scoring methods: cosine similarity and maximal marginal relevance search~\cite{10.1145/290941.291025,langchain}. 
The corpus to use for retrieval is determined by the release number mentioned in the user query (as extracted in Step~1). If no release number is provided, the latest corpus is used to generate responses. Since there are multiple standalone queries, multiple sets of search chunks will be retrieved. 
We take the (duplicate-free) union of these search chunks. Next, using the traceability information retained between search and context chunks in Algorithm~\ref{alg:dual-chunking}, we map the search chunks to their corresponding context chunks. Finally, we proceed to Step~3, described below, with a (duplicate-free) set of the most relevant context chunks, discarding any redundancies.

\textit{\textbf{Step 3: Context Reduction.}} As noted earlier, large context chunks, even when within an LLM's token limit,  can create a  ``needle in a haystack''  problem, where irrelevant details obscure key information and hinder accurate answer generation by LLMs~\cite{hassani2024rethinkinglegalcomplianceautomation,DBLP:conf/emnlp/LabanFXW24,liu2023lostmiddlelanguagemodels,10.1145/3644815.3644945}. To address this, \approach\ uses few-shot, chain-of-thought (CoT) prompts~\cite{qamrprompts} to extract only the information relevant to the user query from each retrieved context chunk. This ensures that each \emph{individual} chunk contains only query-relevant information. Each reduced chunk is retained unless it is empty, indicating that the original chunk contained no relevant information.

\textit{\textbf{Step 4: Context Selection.}}  \approach\ identifies the most relevant reduced context chunks from Step~3 based on their usefulness in answering the corresponding standalone query.
Similar to step 3, this is achieved using a few-shot CoT prompt~\cite{qamrprompts}. The prompt we use here is  inspired by the one proposed by Yu et al.~\cite{yu2024rankragunifyingcontextranking}. 
The top-ranked reduced context chunks are then selected.

\textit{\textbf{Step 5: Answer Generation.}}  \approach\ uses the top-ranked reduced context chunks from Step~4 to answer the query. The prompt~\cite{qamrprompts} instructs \llm\ to rely only on these chunks  rather than its internal knowledge and to reply with ``I don’t know'' if they lack sufficient information for answering the query, reducing hallucination risk.

%% file: Sections/alg.tex
\small
\begin{algorithm}[t]
\caption{\approach's Dual-Chunking Algorithm}
\label{alg:dual-chunking}
{\scriptsize
\begin{flushleft}
\textbf{Input} $\mathit{Docs}$: A set of processed documents from Step~1 of Figure~\ref{figure:corpus_creation}\\
\textbf{Parameter} $\mathit{k}$: Number of search chunks  per page\\
\textbf{Parameter} $\mathit{ps}$: Padding size\\[.5em]
\textbf{Output} $\textit{S-Chunks}$: A set of search chunks\\
\textbf{Output} $\textit{C-Chunks}$: A set of context chunks\\
\end{flushleft}
\begin{algorithmic}[1]
\State $\textit{C-Chunks} = \textit{S-Chunks} \gets \emptyset$
\For{every  $\mathit{doc}_i \in \mathit{Docs}$}
    \For{every  $\mathit{pg}_{ij} \in \mathit{doc}_i$}  \Comment{Building search chunks}
    \State Divide  $\mathit{pg}_{ij}$ into  $k$ search chunks $\mathit{sc}_{ij1}, \ldots, \mathit{sc}_{ijk}$ with equal size  
    \State $\textit{S-Chunks} \gets \textit{S-Chunks} \cup \{\mathit{sc}_{ij1}, \ldots, \mathit{sc}_{ijk}\}$ 
    \EndFor
    \For{every  $\mathit{pg}_{ij} \in \mathit{doc}_i$} \Comment{Building context chunks}
     \State $\mathit{cc}_{ij} \gets \mathit{pg}_{ij}$ \Comment{Initialize context chunk $\mathit{cc}_{ij}$}
    \State  Add $\mathit{ps}$ characters from the bottom of $\mathit{pg}_{i,j-1}$ to  the beginning of $\mathit{cc}_{ij}$ 
    \State  Add $\mathit{ps}$ characters from the top of $\mathit{pg}_{i,j+1}$  to the end of $\mathit{cc}_{ij}$
    \State $\mathit{metadata}(cc_{ij})$ $\gets$  the title info of $\mathit{doc}_{i}$ 
    \State $\textit{C-Chunks} \gets \textit{C-Chunks} \cup \{ \mathit{cc}_{ij} \}$
    \EndFor\ \Comment{Each search chunk $\mathit{sc}_{ijl}$ is related to context chunk $\mathit{cc}_{ij}$.}
\EndFor
\State \Return $\textit{S-Chunks}$, $\textit{C-Chunks}$  \Comment{Returning search and context chunks}
\end{algorithmic}}
\end{algorithm}
\normalsize

%% file: Sections/evaluation.tex
\section{Empirical Evaluation}
\label{sec:eval}
Our evaluation aims to answer three research questions:

\textbf{RQ1 (Accuracy).} \emph{How accurate is \approach?} RQ1 assesses the accuracy of \approach\ using  a proprietary dataset from Ciena and an open-source dataset, REQuestA~\cite{ezzini2023aibasedquestionansweringassistance}.
We compare \approach\ with the baseline described in Section~\ref{sec:baseline} using six LLM-as-Judge metrics~\cite{deepeval,liu2023gevalnlgevaluationusing}, which measure  answer accuracy and  retrieval accuracy. To ensure the reliability of these LLM-generated metrics, we conduct a correlation analysis validating the metrics against expert human judgments.

\textbf{RQ2 (Ablation Study).} \emph{How does each step of   \approach\ affect  its overall accuracy?} Among the five steps of \approach, the last step -- answer generation -- is essential for any question-answering chatbot. The other four steps, however, are designed to improve accuracy, particularly for multi-release systems. RQ2 investigates how each of these four steps individually contributes to the overall accuracy of \approach.

\textbf{RQ3 (Response Time).} \emph{What is the execution time of \approach?} RQ3 compares the response times of \approach\ and the baseline across the two datasets  in our evaluation.

\subsection{Datasets and Ground Truths}
\label{sec:datsets}
Our evaluation uses two datasets: Ciena’s system-specification documents on optical networking and telecommunications, and REQuestA, a public dataset of aerospace, defence, and security software requirements~\cite{ezzini2023aibasedquestionansweringassistance}.

The Ciena dataset -- the larger of the two datasets -- has multi-release characteristics, which are absent in REQuestA. The inclusion of REQuestA in our evaluation serves three purposes: (1) to confirm that \approach\ generalizes beyond the Ciena dataset and maintains accuracy across different types of documents, (2) to systematically evaluate the effectiveness of dual chunking irrespective of the presence of multi-release characteristics, and (3) to be able to provide a public replication package since the Ciena dataset, due to its proprietary nature, cannot be publicly released.

Table~\ref{table:dataset} provides summary statistics for the two datasets. Ciena’s dataset includes 59 multi-release system specifications, averaging 531 pages and 887,559 characters, with 5,694 non-textual elements. In contrast, REQuestA dataset comprises six software requirements specifications, averaging 65 pages and 119,741 characters, with only 20 non-textual elements.

\begin{table}[t]
\centering
\caption{Summary statistics for datasets.}\label{table:dataset}
\scalebox{0.85}{\begin{tabular}{l|cc|}
\cline{2-3}
                                           & \multicolumn{1}{c|}{\textbf{Ciena}} & \textbf{REQuestA} \\ \hline
\multicolumn{1}{|l|}{\textbf{Number of Documents}}      & \multicolumn{1}{c|}{59}             & 6            \\ \hline
\multicolumn{1}{|l|}{\textbf{Average Characters per Document}}  & \multicolumn{1}{c|}{887,559}        & 119,741      \\ \hline
\multicolumn{1}{|l|}{\textbf{Average Pages per Document}}  & \multicolumn{1}{c|}{531}            & 65           \\ \hline
\multicolumn{1}{|l|}{\textbf{Number of Questions}} & \multicolumn{1}{c|}{88}             & 159          \\ \hline
\multicolumn{1}{|l|}{\textbf{Average Characters per User Queries}} & \multicolumn{1}{c|}{74}             & 65          \\ \hline
\multicolumn{1}{|l|}{\textbf{Average Characters per Ground Truth Response}} & \multicolumn{1}{c|}{240}             & 528          \\ \hline
\multicolumn{1}{|l|}{\textbf{Number of Non-Textual Elements}} & \multicolumn{1}{c|}{5694}             & 20          \\ \hline

\end{tabular}}
\vspace*{-0.2cm}
\end{table}

The Ciena documents are accompanied by a ground truth including 88 question-answer pairs, with questions averaging 74 characters and answers averaging 240 characters. This dataset is part of an internal benchmark developed by the company to evaluate chatbot initiatives. Both the questions and answers were carefully created by subject-matter experts, none of whom are co-authors of this paper. The REQuestA ground truth consists of 159 well-defined, human-verified question-answer pairs~\cite{ezzini2023aibasedquestionansweringassistance}, with questions averaging 65 characters and answers averaging 528 characters.

\subsection{Evaluation Metrics} 
\label{sec:metrics}
We evaluate \approach\ using the following six LLM-as-Judge metrics: \emph{answer correctness}, \emph{answer relevancy}, \emph{answer faithfulness}, \emph{contextual precision}, \emph{contextual recall}, and \emph{contextual relevancy}. \emph{Answer correctness} is measured with G-Eval~\cite{liu2023gevalnlgevaluationusing}, while the other five use DeepEval~\cite{deepeval}. All metrics rely on an LLM as the evaluator. We describe these six metrics below:

Let $Q$ be the set of user queries in our evaluation, $R$ the set of generated answers, and $G$ the set of ground-truth answers for $Q$. For each query $q \in Q$, let $r_q \in R$ denote the generated answer and $g_q \in G$ the ground-truth answer. Further, let $C_q = (c_1, \ldots, c_l)$ represent the ordered list of chunks used to generate the answer for $q$. 
We refer to the underlying LLM used for computing the accuracy metrics as \llmEval. Throughout the evaluation, $q$ always refers to the original user query before query rewriting.

\subsubsection{Answer Correctness}
\label{sec:answer_correctnes_metric}
This metric uses \llmEval\ and relies on domain-specific prompts that emulate the reasoning a human evaluator would apply to assess the quality of responses generated within the chatbot's application context~\cite{liu2023gevalnlgevaluationusing}. In collaboration with Ciena, we developed eight prompts, provided online~\cite{qamrprompts}, designed to replicate expert reasoning. These prompts are organized into four groups:

\textbf{Factual Accuracy (two prompts)} ensure that, $r_q \in R$ and $g_q \in G$ are factually consistent for each question $q \in Q$.

\textbf{Matching Units and Values (two prompts)} ensure no mismatch in values or units between $r_q$ and $g_q$, e.g., seconds used instead of decibel-milliwatts for the receiver-damage~level.

\textbf{Equivalence of Core Meaning (two prompts)} ensure that differences in verbosity or in the order of information presented in $r_q$ and $g_q$ are disregarded, as long as $r_q$ and $g_q$ convey the same core meaning. For example, consider a question $q$ asking which pins a specific product can connect to. Suppose $g_q$ explicitly lists the pin numbers, while $r_q$ offers guidelines for determining these pins, such that engineers can identify them using those guidelines. In this example, our prompts ensure that $r_q$ is deemed correct.

\textbf{Recognition of Scope (two prompts)} ensure that: (1) For out-of-scope queries (where $g_q$ indicates no answer exists), $r_q$ explicitly acknowledges the absence of an answer; and (2) For in-scope queries,  $r_q$ does not claim that there is no answer.

Given a set of domain-specific prompts that produce verdicts of ``correct'' or ``incorrect'', G-Eval assigns a score (between 0 and 1) to each query $q$, evaluating $r_q$ against $g_q$~\cite{liu2023gevalnlgevaluationusing}.

\subsubsection{Answer Relevancy} For a query $q$, answer relevancy uses \llmEval\ to classify the statements in $q$'s generated response, $r_q$, as  relevant or irrelevant to the statements in $q$. Let $A$ be the total number of statements in $r_q$, and $B$ be the number of statements in $r_q$ that are relevant to $q$. Answer relevancy is  defined as $\frac{B}{A}$, measuring the proportion of answer \hbox{statements that are relevant to $q$.}

\subsubsection{Answer Faithfulness} For a query $q$,  answer faithfulness
uses \llmEval\   to classify statements in $q$'s generated response, $r_q$, as being relevant or irrelevant to at least one statement in some chunk in $C_q = (c_1, \ldots, c_l)$ used for answering $q$. Let $A$ be the total number of statements in $r_q$, and $B$ be the number of statements in $r_q$ that find a relevant statement in $C_q$. Answer Faithfulness is then defined as $\frac{B}{A}$, which measures the proportion of answer statements that are relevant to the chunks used to answer $q$.

\subsubsection{Contextual Precision} 
For a query $q$, contextual precision measures the relevance of the chunks $C_q = (c_1, \ldots, c_l)$ used for answering it. This metric uses \llmEval\ to classify each chunk $c_i$ as being relevant or irrelevant to $q$ by comparing $c_i$ with $q$'s ground-truth answer, i.e., $g_q$. For each $c_i$, let $b_i$ be a binary variable that is set to 1 when $c_i$ is relevant to $g_q$ and to 0 otherwise. Contextual precision for $q$ is then defined~as:

\hspace*{0.1cm}

\scalebox{0.99}{$\begin{array}{c}
\text{Contextual precision}  =  \frac{  \sum_{i=1}^{l} \left( \frac{\text{\# of relevant chunks in sublist} (c_1, \ldots, c_i)}{i} \times b_i \right) }{\text{\# of relevant chunks in}\ C_q} \\
\end{array}$}\\

Contextual precision emphasizes the relevance of higher-ranked contexts (near~1) more than lower-ranked ones (near~$l$).

\subsubsection{Contextual Recall}  For a query $q$, contextual recall uses \llmEval\ to classify the statements in $q$'s ground-truth answer, $g_q$, as relevant or irrelevant to the chunks $C_q = (c_1, \ldots, c_l)$ used for answering $q$. Let $A$ be the total number of statements in $g_q$, and let $B$  be the number of statements in $g_q$ that are relevant to some statement in some chunk $c_i$. Contextual recall is defined as $\frac{B}{A}$, measuring the proportion of ground-truth statements covered by $C_q$.

\subsubsection{Contextual Relevancy} For a query \( q \), contextual relevancy uses \llmEval\ to classify the statements in chunks \( C_q = (c_1, \ldots, c_l) \) used for answering \( q \) as being relevant or irrelevant to \( q \). Let \( A \) be the total number of statements in \( C_q \), and let \( B \) be the number of statements in \( C_q \) that are relevant to \( q \). Contextual relevancy is then defined as \( \frac{B}{A} \), measuring the proportion of chunk statements that are relevant to \( q \).

\subsection{Ablations and Baseline}
\label{sec:baseline}
We develop five ablations of \approach, and in addition,  compare \approach\  with a state-of-the-art baseline. The five ablations correspond to  four  steps of \approach\ -- query rewriting, dual chunking, context reduction, and context selection -- as well as one ablation that excludes all these steps. Specifically,  we construct (i) one ablation without any of the four intermediate steps but with answer generation, and (ii) four ablations where each includes only a single intermediate step along with answer generation. These ablations enable us to isolate the impact of each component of \approach. In particular, the ablation with query rewriting allows us to assess the effectiveness of this step in improving answer correctness for multi-release documents.

For comparison with existing methods,  we use the architecture of state-of-the-art RAG-based question-answering chatbots in software engineering~\cite{10.1145/3644815.3644945, chaudhary2024developingllamabasedchatbotcicd, ma2023queryrewritingretrievalaugmentedlarge,yu2024rankragunifyingcontextranking}, adapting them to include the query rewriting and context selection steps from \approach. This adaptation enables the baseline to handle multi-release documents, noting that existing chatbots are not suitable for use in their current form with such documents. Our baseline implementation is provided online~\cite{replicationPackage}.

\subsection{Implementation}
For both \approach\ and the baseline, we use Meta's instruction-tuned Llama 3.0 model with 70B parameters (Llama-3-70B-Instruct)~\cite{llama3modelcard}. Llama 3-70B balances efficiency with strong instruction-following capabilities while remaining practical for local deployment~\cite{llama3modelcard} -- as necessitated by our industry context. We implement \approach\ and the baseline using Python 3.10 and the Transformers library (v4.45.1)~\cite{wolf2020huggingfacestransformersstateoftheartnatural} for model loading and language tokenization. To minimize computational overhead, both \llm\ and \llmEval\ are loaded in a 4-bit quantized format using BitsAndBytes (v0.43.1)~\cite{hf_bits_and_bytes}. For document indexing, we use the BAAI/bge-base-en-v1.5 model, which provides high-quality and efficient semantic embeddings, making it ideal for large corpora~\cite{bge_embedding, muennighoff2022mteb}. The dual-chunking approach is implemented using LangChain's Multi-Vector Retrievers~\cite{langchain_multivectorretriever_docs}, with vectors stored in a Chroma vectorstore~\cite{chroma}. We use LangChain (v0.2.8)~\cite{langchain} as the primary underlying framework, utilizing the LangChain Expression Language (LCEL)~\cite{lcel} to construct chatbot pipelines. Our implementation is available online~\cite{replicationPackage}.

\subsection{Experimental Procedure}
\label{subsec:expproc}
Our experiments were conducted on a machine with two Intel Xeon Gold 6338 CPUs, 512~GB of RAM, and one NVIDIA A40 GPU (46 GB memory). To mitigate randomness, each experiment was repeated \emph{ten times}. In total, we posed 4,940 queries across both datasets and evaluated the responses using the six metrics described in Section~\ref{sec:metrics}. 

Table~\ref{table:ds-params} shows the parameters for configuring \approach\ and the baseline. \approach's dual-chunker (Algorithm~\ref{alg:dual-chunking}) requires two parameters: $\mathit{k}$ and $\mathit{ps}$. We set $\mathit{k} = 2$ to split each page into two search chunks, balancing search accuracy and effectiveness by avoiding chunks that are overly large or small. For the padding size $\mathit{ps}$, we use 500 characters to provide sufficient preceding and succeeding context (roughly 1/3 to 1/4 of a page). For the baseline, which uses  the same chunks  for both search and answer generation, we treat each page as a single chunk, with pages exceeding 3,000 characters split into smaller chunks.
The chunks maintain a 25\% overlap with the previous and next chunks, following recommendations from prior studies on traditional (single-chunk) \hbox{RAG systems~\cite{chaudhary2024developingllamabasedchatbotcicd, ezzini2023aibasedquestionansweringassistance}.}

\begin{table}[t]
\centering
\caption{Key parameters of \approach\ and the baseline.}\label{table:ds-params}
\scalebox{0.62}{
\begin{tabular}{
    >{\centering\arraybackslash}p{1.2cm}| 
    >{\centering\arraybackslash}p{2.7cm} 
    >{\centering\arraybackslash}p{2.5cm}| 
    >{\centering\arraybackslash}p{2.5cm} 
    >{\centering\arraybackslash}p{2.5cm}
}
\toprule
 & \multicolumn{2}{c}{\textbf{Chunking}} 
 & \multicolumn{2}{|c}{\textbf{Retrieval}} \\
\midrule
\multirow{4}{*}{\textbf{\approach}} 
  & \textit{\textbf{\# of search chunks per page (k)}} 
 & \textit{\textbf{Padding size (ps)}} 
 & \textit{\textbf{\# of retrieved chunks per standalone query}} 
 & \textit{\textbf{\# of chunks used for answer generation}} \\
 \cmidrule(lr){2-3}
 \cmidrule(lr){4-5}
 & 2 & 500 characters & \multirow{4}{*}{4} & \multirow{4}{*}{3} \\
\cmidrule(lr){1-3}
\multirow{3}{*}{\textbf{Baseline}} 
 & \multicolumn{2}{c|}{\textit{\textbf{Chunk size}}} & & \\
\cmidrule(lr){2-3}
 & \multicolumn{2}{p{6.0cm}|}{Each page up to 3000 characters is a chunk} & & \\
\bottomrule
\end{tabular}
}
\vspace*{-.2cm}
\end{table}

As shown in Table~\ref{table:ds-params}, both \approach\ and the baseline retrieve four chunks per standalone query; two via cosine similarity, and the other two via maximal marginal relevance. The answer generation step then uses the top three ranked chunks to produce responses, consistent with prior work~\cite{chaudhary2024developingllamabasedchatbotcicd, ezzini2023aibasedquestionansweringassistance}.

We evaluate the five ablations discussed in Section~\ref{sec:baseline} using the same setup for \approach\ described in Table~\ref{table:ds-params} on the Ciena dataset and repeat each experiment \emph{ten times}.

\begin{figure}
  \begin{subfigure}{1.05\columnwidth}
     \hspace*{-1em}
     \includegraphics[width=0.9\columnwidth]{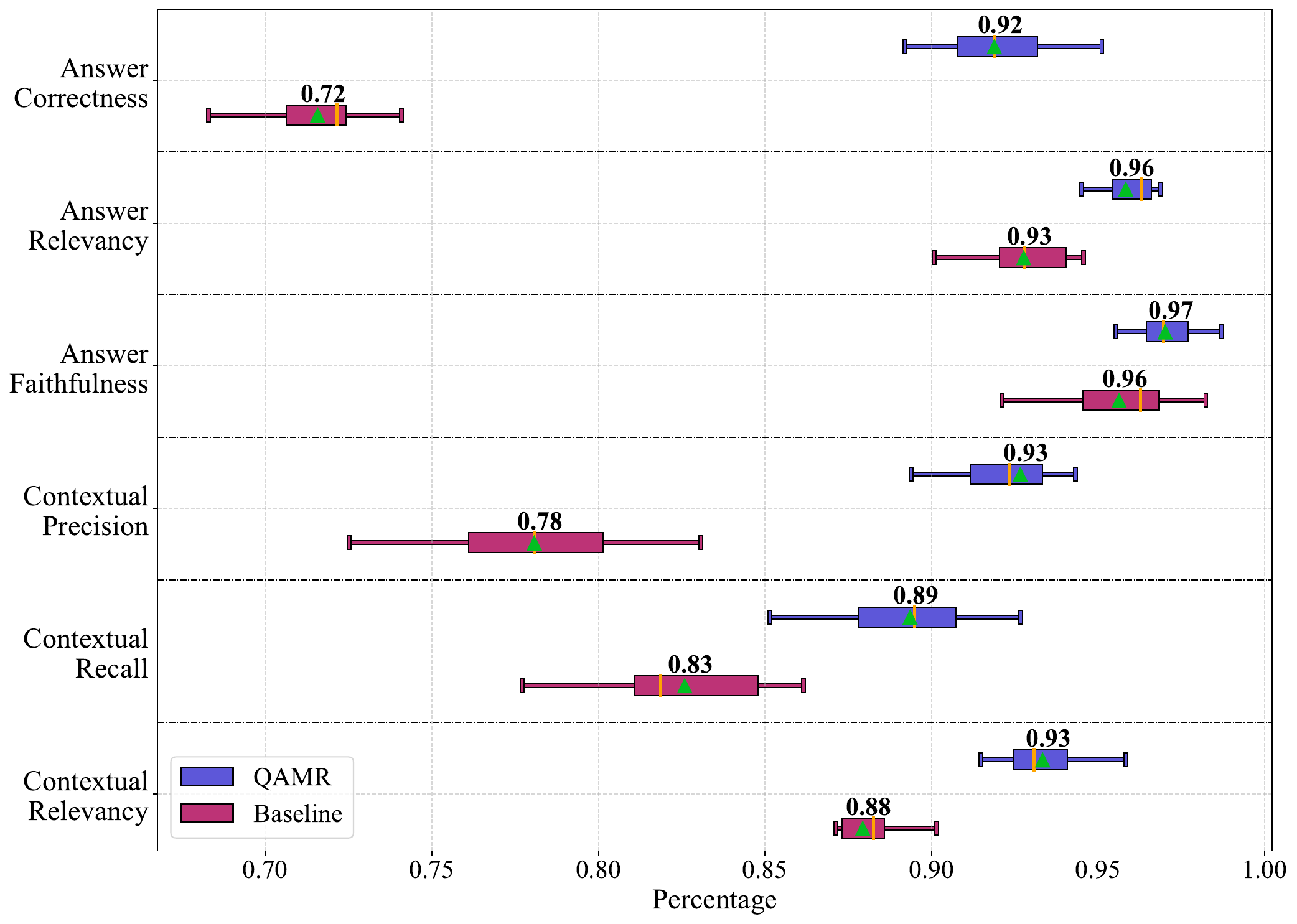}
     \caption{Ciena Dataset}
     \label{figure:ciena_metrics_results}
     \end{subfigure}
     \vspace*{1em}
     \begin{subfigure}{1.05\columnwidth}
   \hspace*{-1em}
    \includegraphics[width=0.9\columnwidth]{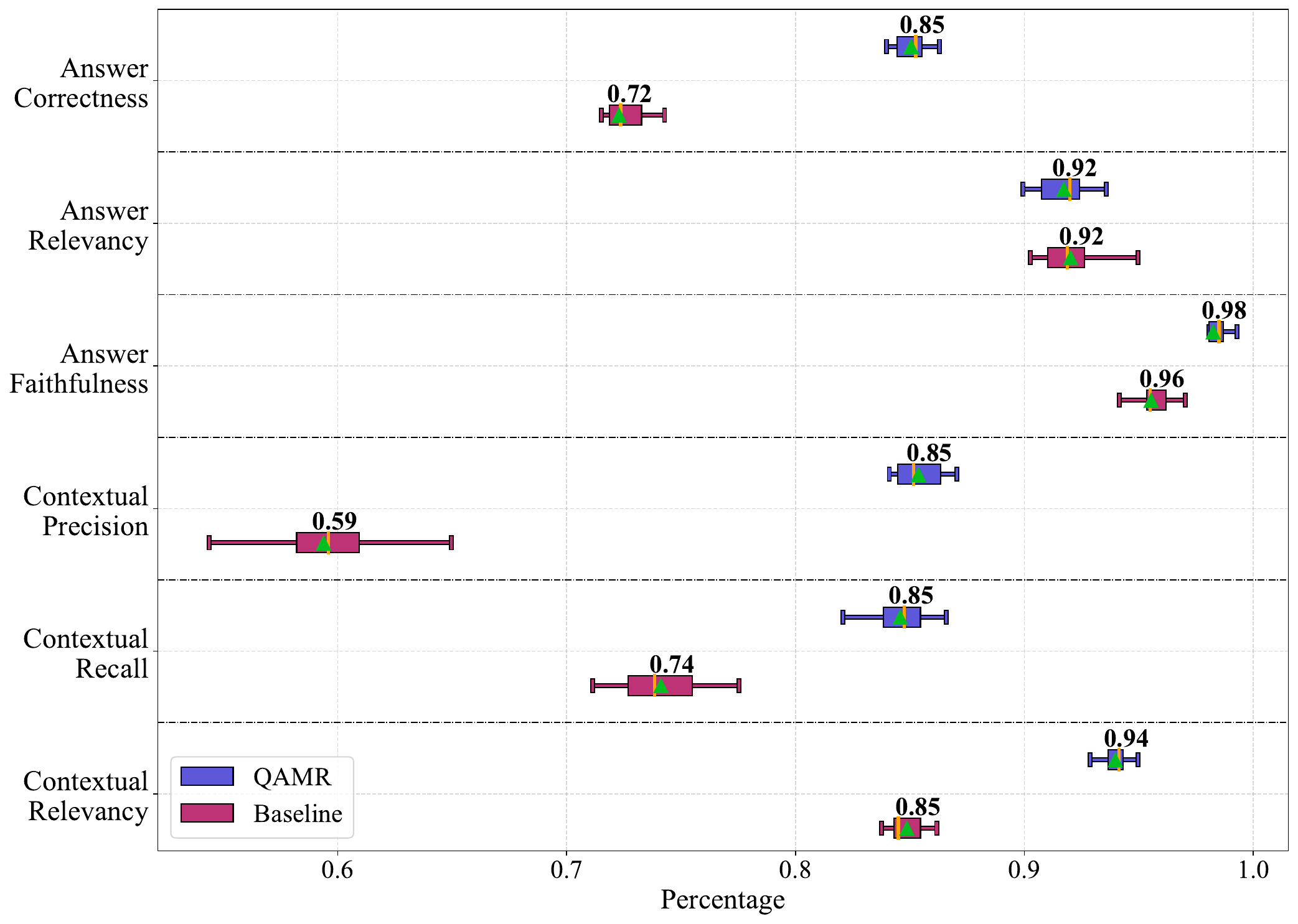}
     \caption{REQuestA Dataset}
     \label{figure:public_metrics_results}
     \end{subfigure}
      \vspace*{-1em}
            \vspace*{-.4cm}
        \caption{Comparison of \approach\ and the baseline using the  metrics defined in Section~\ref{sec:metrics}.}
    \label{fig:allresults}
    \vspace*{-.2cm}
 \end{figure}

\subsection{Results}
In this section, we present our findings and answer RQ1-RQ3. To provide statistical support for our analysis, we use Wilcoxon Rank-Sum test~\cite{wilcoxon1992individual} and Vargha–Delaney effect size ($\hat{A}_{12}$)~\cite{vargha2000critique}, with a 1\% significance level. A difference is considered statistically significant when the $p$-value falls below this threshold.  The effect size is classified as small, medium, or large when $\hat{A}_{12}$ deviates from 0.5 by at least 
0.06, 0.14, and 0.21, respectively.

\textbf{RQ1 (Accuracy).} 
We measure accuracy using the six metrics defined in Section~\ref{sec:metrics}. Figure~\ref{fig:allresults} compares the distributions of the six metrics obtained by \approach\ and the baseline for the Ciena and REQuestA datasets. 
For the Ciena dataset (Figure~\ref{fig:allresults}(a)), \approach\ achieves higher averages than the baseline across all six metrics. For the REQuestA dataset (Figure~\ref{fig:allresults}(b)), \approach\ achieves higher averages for five of the metrics, with both \approach\ and the baseline producing the same average for answer relevancy. 
Statistical tests in Table~\ref{table:stat_evaluation_results} show that \approach\ significantly outperforms the baseline with a large effect size for all metrics except answer faithfulness in the Ciena dataset and answer relevancy in the REQuestA dataset.

\begin{table}[t]
\centering
    \caption{Statistical test results comparing \approach\ and the baseline. \approach\ significantly outperforms the baseline with large effect sizes, except in the yellow highlighted cells where no statistical significance is observed.}    \label{table:stat_evaluation_results}
    \scalebox{0.85}{\begin{tabular}{l|cc|cc|}
        \cline{2-5}
        \multirow{2}{*}{} & \multicolumn{2}{c|}{\textbf{Ciena}} & \multicolumn{2}{c|}{\textbf{REQuestA}} \\ \cline{2-5} 
         & \multicolumn{1}{c|}{p-value} & \multicolumn{1}{c|}{\raisebox{-0.5ex}{$\hat{A}_{12}$}} & \multicolumn{1}{c|}{p-value} & \raisebox{-0.5ex}{$\hat{A}_{12}$} \\ \hline
        \multicolumn{1}{|l|}{Answer Correctness} & \multicolumn{1}{c|}{0.002} & 1 (L) & \multicolumn{1}{c|}{0.002} & 1 (L) \\ \hline
        \multicolumn{1}{|l|}{Answer Relevancy} & \multicolumn{1}{c|}{0.006} & 0.83 (L)& \multicolumn{1}{c|}{\cellcolor{yellow!25} 0.84} &  -- \\ \hline
        \multicolumn{1}{|l|}{Answer Faithfulness} &  \multicolumn{1}{c|}{\cellcolor{yellow!25}0.08} & -- & \multicolumn{1}{c|}{0.002} & 0.98 (L) \\ \hline
        \multicolumn{1}{|l|}{Contextual Precision} & \multicolumn{1}{c|}{0.002} & 1 (L)  & \multicolumn{1}{c|}{0.002} & 1 (L) \\ \hline
        \multicolumn{1}{|l|}{Contextual Recall} & \multicolumn{1}{c|}{0.004} & 0.98 (L)& \multicolumn{1}{c|}{0.002} & 1 (L) \\ \hline
        \multicolumn{1}{|l|}{Contextual Relevancy} & \multicolumn{1}{c|}{0.002} & 1 (L) & \multicolumn{1}{c|}{0.002} & 1 (L) \\ \hline
            \end{tabular}}
\end{table}

Across both datasets, \approach\ achieves average scores of 
$88.5\%$ in answer correctness, $89\%$ in contextual precision, 
$87\%$ in contextual recall, and $93.5\%$ in contextual relevancy. 
These results indicate an average improvement of $16.5\%$ 
in answer correctness compared to the baseline. Averaging the improvements across contextual precision, recall, and relevancy yields a $12$\% overall increase over the baseline.

As noted in Section~\ref{sec:intro}, we use LLM-as-Judge metrics instead of expert judgment because of the scale of our evaluation and the infeasibility of comprehensive manual assessment. An important question that arises here is how closely our LLM-as-Judge metrics align with expert judgment. To this end, we asked Ciena experts (who are not authors of this study) to review the responses generated by \approach\ on the full Ciena dataset (all 88 questions). From the 10 runs conducted for each question in this dataset, we randomly selected one run, and experts labelled its answer as ``adequate'' or ``inadequate'' based on how closely it matched the ground truth. Similarly, we randomly selected one run of \approach\ on the REQuestA dataset and engaged an independent annotator (also not an author) to label the responses generated for the 159 questions in that dataset, in order to determine how closely the generated answers matched the ground truth. We then computed the Pearson correlation coefficients between the labels provided by humans and the LLM-as-Judge metric values, as reported in Table~\ref{tab:correlations}. In this analysis, among the six metrics discussed in Section~\ref{sec:metrics}, we considered only the metrics that compare generated answers with ground truths -- namely, answer correctness, contextual precision, and contextual recall.  As shown in the table, answer correctness strongly correlates with human judgments (0.88 for Ciena and 0.75 for REQuestA, \hbox{p-value $\ll$ 0.01}). For REQuestA, contextual precision and  recall also correlate positively with human judgment.

\begin{table}[t]
\centering
\caption{Pearson correlation coefficients between  metrics  in Section~\ref{sec:metrics} and adequacy labels provided by domain experts. Green cells represent p-values $\mathbf{ \ll 0.01}$.}
\label{tab:correlations}
\scriptsize
\begin{tabular}{llc}
\toprule
\textbf{Dataset} & \textbf{Metric} & \textbf{Correlation} \\ \midrule
\multirow{3}{*}{Ciena}    
  & Answer Correctness   & \cellcolor{green!20}0.88 \\
  & Contextual Precision & -0.02 \\
  & Contextual Recall    & -0.01 \\ \midrule
\multirow{3}{*}{REQuestA} 
  & Answer Correctness   & \cellcolor{green!20}0.75 \\
  & Contextual Precision & \cellcolor{green!20}0.63 \\
  & Contextual Recall    & \cellcolor{green!20}0.54 \\ 
\bottomrule
\end{tabular}
\vspace*{-0.2cm}
\end{table}

\vspace*{-.2cm}
\begin{tcolorbox}[breakable,colback=gray!10!white, colframe=black!75!black, boxrule=0.5mm, arc=1mm, left=1mm, right=1mm, top=1mm, bottom=1mm, fonttitle=\bfseries]
\textbf{Finding 1.} \approach\ significantly outperforms the baseline on all answer accuracy and retrieval accuracy metrics, except for answer faithfulness (Ciena dataset) and answer relevancy (REQuestA dataset). Notably, \approach\ achieves an average 16.5\% increase in answer correctness and an average improvement of 12\% in contextual precision, recall, and relevance compared to the baseline. \\
\textbf{Finding 2.}  The answer-correctness metric has a strong positive correlation with human assessments for both datasets. \\
\textbf{Take away 1.} \approach's dual chunking identifies relevant chunks more often than traditional chunking. Combined with context reduction, this ensures that answer generation is provided with only relevant information. \\
\textbf{Take away 2.}  The strong correlation between the  answer-correctness metric and human judgment across both datasets indicates that,  in our application context,  this metric is a reliable proxy and provides a scalable alternative to human evaluation for verifying LLM-generated texts against the ground truth.
\end{tcolorbox}

\vspace*{-0.1cm}
\textbf{RQ2 (Ablation Study).} 
Table~\ref{table:ablation-study} presents the ablation study results obtained using  the Ciena dataset. We refer to the ablation that uses only the  answer-generation step as the \emph{base} ablation. Each of the other four ablations combines the base with exactly one of the following \approach's steps: query rewriting, dual chunking, context reduction, or context selection. Box plots showing the distributions underlying the averages in Table~\ref{table:ablation-study} are provided in the replication package~\cite{ablationresults}.

\begin{table}[t]
    \caption{Average (\%) evaluation results for the ablation study. 
The \emph{Base} ablation uses only the answer generation step of \approach. Highlighted cells mark the best-performing ablation for each metric.}
    \label{table:ablation-study}

    \centering
    \scriptsize
    \begin{tabular}{p{2.7cm} | c c c c c}
        \toprule
        & \textbf{Base} 
        & \makecell{\textbf{+ Query}\\ \textbf{Rewrite}} 
        & \makecell{\textbf{+ Dual}\\ \textbf{Chunk}} 
        & \makecell{\textbf{+ Ctx}\\ \textbf{Reduce}} 
        & \makecell{\textbf{+ Ctx}\\ \textbf{Select}} \\ 
        \midrule
        \rowcolor{gray!10} Answer Correctness (\%)   & 68.29 & {\cellcolor{green!25}72.28} & 72.13 & 59.30 & 64.39 \\
        Answer Relevancy (\%)     & 93.56 & 92.64 & {\cellcolor{green!25}93.67} & 78.07 & 90.69 \\
        \rowcolor{gray!10} Answer Faithfulness (\%)  & 95.04 & 94.95 & 96.24 & {\cellcolor{green!25}97.91} & 95.09 \\
        Contextual Precision (\%) & 67.46 & 71.25 & 57.77 & 58.88 & {\cellcolor{green!25}72.39} \\
        \rowcolor{gray!10} Contextual Recall (\%)    & 73.26 & {\cellcolor{green!25}74.76} & 63.93 & 61.45 & 73.25 \\
        Contextual Relevancy (\%) & 82.10 & 78.82 & 75.97 & 74.68 & {\cellcolor{green!25}87.63} \\
        \bottomrule
    \end{tabular}
\vspace*{-0.2cm}
\end{table}

The results  in Figure~\ref{fig:allresults}(a) and  Table~\ref{table:ablation-study} show that  \approach\ significantly improves the average answer correctness, answer relevancy, contextual precision, contextual recall, and contextual relevancy relative to the five ablations, with minimum respective improvements of 19.60\%, 2.16\%, 20.27\%, 14.59\%, and 5.70\%. In addition, \approach’s performance on the answer faithfulness metric is better or on par with that of the ablations.

Statistical tests in Table~\ref{table:ablation_evaluation_results} show that \approach\ significantly outperforms all ablations with a large effect size for all metrics except for the answer relevancy and answer faithfulness metrics. For answer relevancy, \approach\ significantly outperforms two ablations, while for answer faithfulness, it significantly outperforms one ablation. In both cases, the differences show large effect sizes. For all other comparisons of answer faithfulness and relevancy, there is no statistically significant difference between \approach\ and its ablations. Thus, according to our statistical tests, \approach\ is never outperformed by any of its ablations on any metric.

\begin{table}[t]
\caption{Statistical test results comparing \approach\ with each of  the  ablations in Table~\ref{table:ablation_evaluation_results}. 
\approach\ significantly outperforms all ablations with large effect sizes, 
except in the yellow highlighted cells where no statistical significance is observed.}
\label{table:ablation_evaluation_results}
\resizebox{\linewidth}{!}{
\begin{tabular}{lcccccccccc}
\toprule
 & \multicolumn{2}{c}{\textbf{Base}} 
 & \multicolumn{2}{c}{\textbf{+ Query Rewrite}} 
 & \multicolumn{2}{c}{\textbf{+ Dual Chunk}} 
 & \multicolumn{2}{c}{\textbf{+ Ctx Reduce}} 
 & \multicolumn{2}{c}{\textbf{+ Ctx Select}} \\
\cmidrule(lr){2-3}\cmidrule(lr){4-5}\cmidrule(lr){6-7}\cmidrule(lr){8-9}\cmidrule(lr){10-11}
 & p-value & $\hat{A}_{12}$ & p-value & $\hat{A}_{12}$ & p-value & $\hat{A}_{12}$ & p-value & $\hat{A}_{12}$ & p-value & $\hat{A}_{12}$ \\
\midrule
Ans. Cor.      & 0.002 & 1(L) & 0.002 & 1(L) & 0.002 & 1(L) & 0.002 & 1(L) & 0.002 & 1(L) \\
Ans. Rel.     & \cellcolor{yellow!25}0.064 & - & \cellcolor{yellow!25}0.080 & - & 0.002 & 0.75(L) & 0.002 & 1(L) & \cellcolor{yellow!25}0.020 & - \\
Ans. Fait.  & \cellcolor{yellow!25}0.064 & - & 0.002 & 0.84(L) & \cellcolor{yellow!25}0.080 & - & \cellcolor{yellow!25}0.200 & - & \cellcolor{yellow!25}0.014 & - \\
Ctx. Prec.     & 0.002 & 1(L) & 0.002 & 1(L) & 0.002 & 1(L) & 0.002 & 1(L) & 0.002 & 1(L) \\
Ctx. Rec.        & 0.002 & 1(L) & 0.002 & 1(L) & 0.002 & 1(L) & 0.002 & 1(L) & 0.002 & 1(L) \\
Ctx. Rel.     & 0.002 & 1(L) & 0.002 & 1(L) & 0.002 & 1(L) & 0.002 & 1(L) & 0.002 & 1(L) \\
\bottomrule
\end{tabular}
}
\end{table}

When comparing the base ablation with the four non-base ablations --~ each incorporating a single step from \approach~-- we observe that each non-base ablation outperforms the base ablation in at least one metric. Specifically, the query-rewriting ablation improves answer correctness, answer faithfulness, contextual precision, and contextual recall compared to the base ablation; the dual-chunking ablation improves answer correctness, answer faithfulness, and answer relevancy; the context-selection ablation improves contextual precision, contextual recall, and contextual relevancy; and the context-reduction ablation improves answer faithfulness. Notably, the base ablation never achieves the highest average on any metric.

\begin{tcolorbox}[breakable,colback=gray!10!white, colframe=black!75!black, boxrule=0.5mm, arc=1mm, left=1mm, right=1mm, top=1mm, bottom=1mm, fonttitle=\bfseries]
\textbf{Finding 1.} No ablation significantly outperforms \approach\ in any metric, while \approach\ significantly outperforms all ablations in four or five out of the six metrics. Further, every non-base ablation outperforms the base in at least one metric, showing that each step -- query rewriting, dual chunking, context reduction, and context selection -- contributes positively to the overall approach. All steps together complement one another, enabling \approach\ to outperform every ablation. \\
\textbf{Finding 2.} The query-rewriting ablation outperforms the other ablations that lack mechanisms for interpreting queries across multi-release documents. Among the five ablations (Table~\ref{table:ablation-study}), the query-rewriting ablation achieves the highest answer correctness and contextual recall, as well as the second-highest contextual precision. This highlights the importance of query rewriting for accurate question answering in the context of multi-release documents.
\end{tcolorbox}

\vspace*{-0.1cm}
\textbf{RQ3 (Response Time).} Table~\ref{table:response-time-practical} compares the average response times of \approach\ and the baseline. Both yield higher average response times on the Ciena dataset than on REQuestA, as Ciena's documents are larger and more complex, often containing multiple tables and schematics.  We observe two points from Table~\ref{table:response-time-practical}:  (1)~For both datasets, the average response time of the baseline is about 8\% higher than that of \approach\ (Ciena: 45/41, REQuestA: 40/37). (2)~On the Ciena dataset, the average response times of both \approach\ and the baseline are 12\% higher than on REQuestA (\approach: 41/37, baseline: 45/40), indicating that \approach’s runtime scales proportionally with the baseline as document complexity increases.

As shown in Table~\ref{table:response-time-practical}, for \approach, context reduction is the most time-consuming step, accounting for 36\% of the total response time, followed by query rewriting and context selection at 27\% and 23\%, respectively. For the baseline, which does not have a context-reduction step, answer generation is the most time-intensive step, followed by context selection and query rewriting. In the baseline, answer generation and context selection are more expensive than query rewriting, whereas the opposite is true for \approach. This is because, without context reduction, the baseline must process large chunks, while context reduction provides a concise context for answer generation and selection, thus improving efficiency.

\begin{table}[t]
\centering
\caption{Average response times of \approach\ vs. baseline.}
\label{table:response-time-practical}
\scalebox{0.65}{
\begin{tabular}{l|cc|p{1.3cm}p{1.45cm}p{1.35cm}p{1.35cm}p{0.65cm}|}
\cline{2-8}
 & \multicolumn{2}{c|}{\bf Average Response Time} & \multicolumn{5}{c|}{\bf Response Time Breakdown} \\
 & \multicolumn{2}{c|}{\bf per Query (in seconds)} & \multicolumn{5}{c|}{\bf according to Steps (\%)} \\ \cline{2-8} 
 & \multirow{2}{*}{Ciena}  & \multirow{2}{*}{REQuestA} & Query Rewriting & Context \linebreak
 Reduction & Context \linebreak Selection & Answer Generation  & Other Steps \\
\hline
\multicolumn{1}{|l|}{\textbf{\approach}} & 41 & 37 & 27\% & 36\% & 23\% & 13\% & 1\% \\ \hline
\multicolumn{1}{|l|}{\textbf{Baseline}} & 45 & 40 & 20\% & N/A & 33\% & 46\% & 1\% \\ \hline
\end{tabular}}
\vspace*{-.2cm}
\end{table}

\color{black}

\begin{tcolorbox}[breakable,colback=gray!10!white, colframe=black!75!black, boxrule=0.5mm, arc=1mm, left=1mm, right=1mm, top=1mm, bottom=1mm, fonttitle=\bfseries]
\textbf{Finding.} On average, \approach\ is about 8\% faster than the baseline on both datasets, primarily because it provides the answer generation and context selection steps with a more concise context from the context-reduction step. \\
\textbf{Take away.} Based on an internal deployment of \approach\ on Ciena's infrastructure, experts found \approach’s response time to be practical.
\end{tcolorbox}

\subsection{Validity Considerations}
\mbox{}\indent\textbf{Internal Validity.} To ensure a fair comparison, \approach\ and the baseline used the same underlying LLM (\llm), identical prompts for shared components, and the same parameters for retrieval (Table~\ref{table:ds-params}). The baseline's single-chunking approach was configured according to best practices from prior studies~\cite{ezzini2023aibasedquestionansweringassistance,chaudhary2024developingllamabasedchatbotcicd}. Metrics for 
both \approach\ and the baseline were computed using the same evaluation LLM (\llmEval). To mitigate randomness, we set the temperature to 0.01 and top-k to one, and repeated each experiment ten times~\cite{achiam2023gpt,guo2024deepseek}.  Regarding data leakage, we are confident that the proprietary Ciena dataset was not part of \llm's pre-training. While the REQuestA dataset is open-source, both \approach\ and the baseline rely on the same \llm, so any leakage affecting \llm\ impacts both equally. As such, our comparison remains fair and reliable.

\textbf{External Validity.} Our evaluation used two datasets from different sources. The Ciena dataset contains a large collection of complex, multi-release system documents with expert-created and validated questions and ground truths. The REQuestA dataset includes both auto-generated and manual questions, with all auto-generated items rigorously vetted for relevance and quality.
In relation to the choice of LLM, we used Llama 3-70B~\cite{llama3modelcard} for both \llm\ and \llmEval. At the time our experiments were conducted, our industry partner’s policies prohibited use of externally hosted LLMs when handling proprietary data. In addition, our LLM-as-Judge evaluation would have been very costly -- thousands of dollars -- due to the large number of LLM calls required to compute each metric, had it been conducted on externally hosted LLMs.
While benchmarking other LLMs is valuable, the exclusive use of Llama 3-70B does not pose a significant risk, as it remains one of the most accurate instruction-following models -- freely available for research and industry use -- and suitable for on-premise deployment, offering performance competitive with newer LLMs in text generation and question answering~\cite{chiang2024chatbot}.

%% file: Sections/related.tex
\section{Related Work}
\label{subsec:rel}

Prior to LLMs, extractive BERT-based models were used to automate question answering, either by integrating RAG~\cite{ezzini2023aibasedquestionansweringassistance} or without it~\cite{DBLP:conf/cain/BorgBOHGT22}.
More recent approaches rely on generative LLMs such as ChatGPT~\cite{achiam2023gpt} and Llama~\cite{llama3modelcard}, which not only identify relevant text but also generate human-like responses tailored to the user query and conversational history. For example, Chaudhary et al.~\cite{chaudhary2024developingllamabasedchatbotcicd} develop an LLM-based chatbot to help systems engineers with CI/CD-related queries. Similarly, Barnett et al.~\cite{10.1145/3644815.3644945} report on their experience applying LLM-based RAG chatbots to case studies from different domains and provide recommendations for improving such chatbots in the context of software engineering. 
These existing RAG architectures  rely on a traditional single-chunking strategy, similar to our evaluation baseline in Section~\ref{sec:eval}. 

As observed in recent literature~\cite{10.1145/3644815.3644945, chaudhary2024developingllamabasedchatbotcicd, ezzini2023aibasedquestionansweringassistance,ma2023queryrewritingretrievalaugmentedlarge,DBLP:journals/ijmi/LandschaftAMKRWHA24,yu2024rankragunifyingcontextranking,zhu2024largelanguagemodelsinformation}, RAG solutions face important limitations, including challenges in determining the optimal chunk size, the retrieval of irrelevant or noisy data within chunks, and vulnerability to low-quality user queries. Our approach seeks to mitigate these well-known limitations. In particular, our dual-chunking strategy decouples the chunks used for retrieval and generation, thus eliminating the need to select a single chunk size for both tasks. In addition, our approach incorporates context reduction and selection mechanisms to reduce noisy data within chunks. Finally, the query-rewriting component in our approach, which generates multiple standalone queries that improve and extend the user query in various ways, helps improve resilience to low-quality user queries compared to techniques that rewrite the user query into a single standalone query.

In relation to the retrieval step in RAG, Liu et al.~\cite{DBLP:conf/emnlp/LiuHZYXY21} propose a hierarchical  approach that, given a set of documents, first performs document retrieval and then retrieves relevant passages within the retrieved documents. This approach works well when the documents are sufficiently distinct, allowing the document-retrieval  stage to meaningfully narrow the search space. However, this hierarchical method cannot effectively  distinguish between highly overlapping multi-release documents and, similar to the other approaches discussed above,  fails to address the challenges of question answering for multi-release documents described in Section~\ref{sec:intro}.

%% file: Sections/lessons.tex
\section{Lessons Learned}
\label{sec:lessons}
\emph{\textbf{Lesson 1:}} Among the six metrics presented in Section~\ref{sec:metrics}, answer correctness, contextual precision, and contextual recall evaluate generated answers against ground truths, whereas contextual relevancy, answer relevancy, and answer faithfulness do not. Instead, the latter metrics assess how well a response overlaps with the user query or retrieved chunks, rather than whether it matches the ground truth.  High scores on these metrics can be achieved by matching the query or retrieved chunks even when the answer itself is incorrect. In our experiments, this effect made the baseline appear comparable to \approach\ on these three metrics, despite performing substantially worse on the other three metrics that are ground-truth-dependent.

\emph{\textbf{Implication of Lesson~1.}} 
Treat contextual  and answer relevancy, and faithfulness as indicators of plausibility rather than correctness. Interpret them alongside ground-truth-based metrics, and prioritize the latter when assessing accuracy.

\emph{\textbf{Lesson 2:}} In domain-specific and technical documents, even small wording changes can affect chatbot accuracy. For example, asking ``\emph{What are performance monitoring considerations?}'' yields an accurate answer, whereas replacing \emph{``performance monitoring''} with \emph{``PM''} or \emph{``operational''} -- terms not used in the documents -- produces poor results. This is because retrieval is largely literal: if the query vocabulary does not appear in the corpus, relevant chunks are not retrieved.

\emph{\textbf{Implication of Lesson 2:}} Apply query rewriting based on a domain-specific glossary to align user queries with document language by disambiguating and expanding acronyms, and mapping terms to their glossary synonyms.

\emph{\textbf{Lesson 3:}} The most prominent source of error in the baseline, \approach, and its ablations is incorrect document retrieval, as most errors arise from the retrieved context being either irrelevant or incomplete.  The next major source of error is hallucination, which occurs more frequently in the baseline  than in \approach. 
This is because query rewriting improves retrieval accuracy by enabling release-specific retrieval across multi-release documents, while dual-chunking and context reduction further reduce hallucinations by improving retrieval accuracy and pruning irrelevant information.

\emph{\textbf{Implication of Lesson 3.}} 
Apply query rewriting for release-specific retrieval in multi-release systems, dual-chunking to improve overall retrieval accuracy, and context reduction to prune irrelevant information from the retrieved contexts.
\color{black}

%% file: Sections/con.tex
\section{Conclusion}\label{sec:conclusion}
We presented \approach, a RAG-based chatbot for question answering over multi-release systems. We evaluated \approach\ on two datasets, comparing its effectiveness and efficiency against a state-of-the-art baseline and conducting an ablation study to assess the impact of its components, especially the release-specific query-rewriting module.  \approach\ significantly outperforms both the baseline and its ablations in answer correctness and retrieval accuracy. Moreover, the automated LLM-as-Judge metrics used to evaluate answer accuracy closely align with expert evaluations, validating the reliability of these automated metrics as a proxy for manual judgment.

\section{Acknowledgment}
We gratefully acknowledge funding from Mitacs Accelerate, Ciena, the Ontario Graduate Scholarship (OGS) program, and NSERC of Canada under the Discovery and Discovery Accelerator programs.